\documentclass[prl,twocolumn,showpacs,preprintnumbers,amsmath,amssymb]{revtex4-1}
\usepackage{amsfonts}
\usepackage{graphicx}
\usepackage{dcolumn}
\usepackage{float}
\usepackage{bm}

\renewcommand{\O}{ {\cal{O}} }
\newcommand{\com}[1]{}

\renewcommand{\Re}{\mathrm{Re}}

\newcommand{\bS}{\mathbf{S}}
\newcommand{\q}{\mathbf{q}}

\newcommand{\rr}{\mathbf{r}}

\newcommand{\bsigma}{{\boldsymbol\sigma}}

\newcommand{\pdfeps}{eps}

\begin{document}

\title{Chirality waves in two-dimensional magnets}

\author{D. Solenov}
\affiliation{Theoretical Division, Los Alamos National Laboratory, Los Alamos, NM
87545, USA}
\author{D. Mozyrsky}
\affiliation{Theoretical Division, Los Alamos National Laboratory, Los Alamos, NM
87545, USA}
\author{I. Martin}
\affiliation{Theoretical Division, Los Alamos National Laboratory, Los Alamos, NM
87545, USA}

\begin{abstract}
We theoretically show that moderate interaction between electrons confined to move in a plane and localized magnetic moments leads to formation of a noncoplanar magnetic state. The state is similar to the skyrmion crystal recently observed in cubic systems with the Dzyaloshinskii-Moriya interaction; however, it does not require spin-orbit interaction.  The non-coplanar magnetism is accompanied by the ground-state electrical and spin currents, generated via the real-space Berry phase mechanism. We examine the stability of the state with respect to lattice discreteness effects and the magnitude of magnetic exchange interaction. The state can be realized in a number of transition metal and magnetic semiconductor systems.
\end{abstract}


\maketitle

\begin{figure}
\includegraphics[width=0.8\columnwidth]{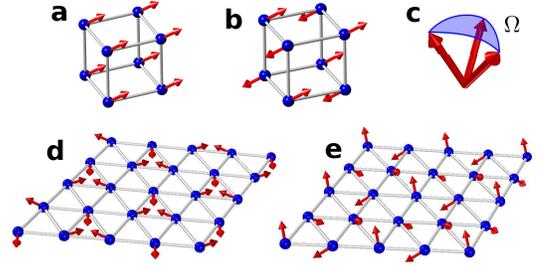}
\caption{Types of magnetic ordering. Collinear magnetic states:
(a) ferromagnet and (b) antiferromagnet. (d) Non-collinear (but
coplanar) magnetic spiral state. (e) Example of {\em
non-coplanar} magnetic state, from Ref.~\cite{Martin}. (c) Solid angle $\Omega$ subtended by three local moments forming an elementary plaquette of a triangular lattice induces a quantum Berry phase $\pm \Omega/2$ for electrons circling the plaquette, locally aligned/antialigned with the texture.
}\label{Fig:mag}
\end{figure}

Magnetism is a cooperative phenomenon where spins of magnetic ions
spontaneously orient relative to each other below certain ordering
temperature \cite{grunberg1986, jaime2001}. In principle, arbitrarily complex magnetic orderings are possible; however, the magnetic states  encountered in
nature tend to be simple, the most common being ferromagnetism
(one atom in magnetic unit cell, Fig.~\ref{Fig:mag}a) and
antiferromagnetism (two distinct atoms in magnetic unit cell,
Fig.~\ref{Fig:mag}b). More complex orders
\cite{Bramwell2001,Braun2005,MNSIYu,BodePappas}, such as
noncollinear spirals (Fig.~\ref{Fig:mag}d) and various noncoplanar
orders (e.g., Fig.~\ref{Fig:mag}e) are less common, typically
arising from the interplay of magnetic exchange interactions,
spin-orbit (SO) coupling \cite{BodePappas}, frustrated lattice
structure \cite{Bramwell2001},  and  magnetic
field \cite{MNSIYu}.

Noncoplanar magnetism has a unique effect on electronic transport
through coherently influencing the quantum-mechanical phase of
electrons  \cite{Ye1999}.  The phase accumulates when an electron moves through the magnetic texture and adjusts its spin according to the
local magnetic environment. The resulting  Berry phase is
equal to half of the solid angle subtended by electron spin in the
process of its evolution around a closed trajectory (Fig.~\ref{Fig:mag}c). This is similar to the Aharonov-Bohm phase induced by the electromagnetic vector potential which couples to electron charge. If the magnetic  texture varies significantly on the scale of the material unit cell, the equivalent magnetic field strength can be gigantic, exceeding 10$^4$ Tesla. Consequently, in non-coplanar magnets one may find such exotic
magneto-transport phenomena as the intrinsic anomalous Hall
 effect \cite{TaguchiNeubauer}, Hall effect in the absence
of net magnetization or magnetic field \cite{Machida2010}, or even { quantum} anomalous Hall effect \cite{OhgushiShindou,Martin}.

In the absence of SO coupling  or magnetic field,
the non-coplanar states  have been believed to be very rare
\cite{Momoi1997, domenge2008}. Recently, however, several examples have been found that  show energetic preference for
non-coplanar magnetic states \cite{Martin, Chern2010,
Akagi2010, Kato2010, Brink2010} in one of the simplest models of
magnetism -- the isotropic Kondo lattice model -- even in the
absence of  SO interaction. These noncoplanar states were found for particular lattice structures and electron densities, which leaves an open question: How
common are the non-coplanar magnetic states? Here, we show that
non-coplanar magnetic states  occur in the
low-density regime of 2D Kondo lattice models without fine tuning.   Our
numerical results suggest that in the weak-to-moderate coupling regime, the lowest energy  state is a ``skyrmion crystal" \cite{Rossler2006}, with the spatial period
determined by the Fermi wavelength,
$\lambda_F \sim 1/q_F$. The texture ${\bf S(r)}$ can be characterized by the scalar spin chirality density, $\kappa = {\bf
S}\cdot [\partial_x {\bf S} \times \partial_y{\bf S}]$, which measures the density of the Berry phase (``Berry
curvature"), and hence is related to the effective orbital magnetic field acting on electrons. We find that $\kappa$ is spatially
modulated,  leading to persistent bulk electrical and spin currents
within the ground state.

The starting point of our analysis is the continuum limit of the Kondo lattice model, which describes  itinerant
electrons interacting with localized moments,
\begin{equation}\label{eq:H}
{\cal H} = -\Psi^\dag\frac{\hbar^2\partial_\rr^2}{2m} \Psi -
J\bS(\rr)\cdot\Psi^\dag\bsigma\Psi \equiv \Psi^\dag\hat H\Psi.
\end{equation}
Here $\Psi = [\psi_\uparrow(\rr), \psi_\downarrow(\rr)]^T$ is the
itinerant electron field operator, $m$ is the electron mass, $\bsigma = (\sigma_x, \sigma_y,\sigma_z)$ is vector of the Pauli matrices and  ${\bf S({\bf
r})}$ is the spin of the magnetic atom located at ${\bf r}$. The
local magnetic moments may originate from the electrons partially
occupying $d$- or $f$-core levels, whose spins are aligned by the
ferromagnetic Hund's interaction. When their combined angular
momentum is large, the behavior of ${\bf S({\bf r})}$ becomes
nearly classical, i.e., slow compared to itinerant electrons. From
the stand point of itinerant electrons the problem then becomes
tractable, since the Hamiltonian is quadratic in electron fields
and therefore can be exactly numerically diagonalized for an arbitrary static
configuration of the classical moments $\bS(\rr)$. The sign of the
Kondo coupling $J$ does not matter in the case of classical local
moments, so we take $J>0$.
\begin{figure}
\includegraphics[width=0.999\columnwidth]{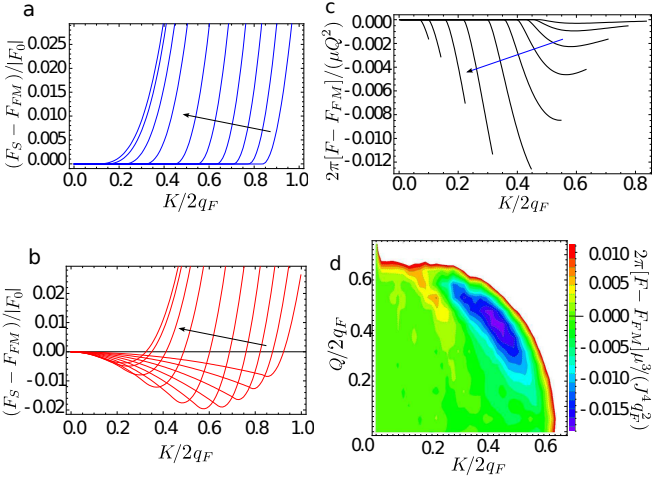}
\caption{Instability of coplanar  spiral state. (a,b)
Free energy of the spiral state $\bS(\rr)=(\sin Ky,0,\cos Ky)$ with
the ordering vector $K$
 for the Kondo coupling values
$J/\mu=0.3,0.4,0.5,0.6,0.7,0.8,0.9,0.95,0.98,0.99$ (arrows
indicate the direction of increasing $J/\mu$). (a) Exact
degeneracy of all spiral states with $K<2q_F\sqrt{1-J/\mu}$. (b)
Lifting of the degeneracy in favor of $K\ne 0$ states due to 
non-parabolicity  of the electron band. Here, the quartic terms
in the dispersion have been introduced  to replicate the effect of
the finite electron bandwidth, $\mu/t=0.24$, with $t$ the nearest
neighbor hopping on a square lattice. (c,d) Energy of the
non-coplanar variational {\em ansatz} $\tilde\bS(\rr)=(\sin Ky\cos
Qx,\sin Ky\sin Qx,\cos Ky)$. (c) Semi-analytical result of the
small-$Q$ expansion for the scaled energy difference $2\pi[F(K,Q)
- F_{FM}]/(\mu Q^2)$, obtained in the continuum limit ($J/\mu$
increases from right to left curves as before). The difference
becomes negative in some range of values $K_0< K <
2q_F\sqrt{1-J/\mu}$, indicating energetic preference for states
with $Q\ne 0$. (d) Free energy of the same {\em ansatz} for
arbitrary values of $Q$ obtained by direct numerical
diagonalization of the Hamiltonian with $J/\mu =
0.67$.}\label{Fig:Q}
\end{figure}

The simplest approach to look for the ordered states  in the Kondo
lattice model is by integrating out the electronic
degrees of freedom perturbatively in  $J$. In
the lowest (2nd) order in $J$, the result is the {classical} Heisenberg
Hamiltonian \cite{RK1954},
\begin{equation}
H_{RKKY} = - J^2 {\sum}_{i,j}\chi({\bf r}_i - {\bf r}_j)\, {\bf
S}_i\cdot {\bf S}_j, \label{eq:Heff}
\end{equation}
that describes a system of  local moments coupled by pairwise
interactions. The spatial dependence of the interaction is
determined by the non-interacting electron spin susceptibility
$\chi({\bf r})$. Due to translational invariance, the ground state
of this approximate model can be easily found by transforming to
the Fourier space \cite{Villain1977}, $H_{RKKY} = - J^2 \sum_{\bf
q}\chi({\bf q})\, {\bf S}_{-\bf q}\cdot {\bf S}_{\bf q}$.  From
the fixed length (classical) constraint, $|\bS_i| = 1$, it follows
that  $\sum_{\bf q} {\bf S}_{-\bf q}\cdot {\bf S}_{\bf q} = N$,
where $N$ is the number of magnetic atoms in the system.
Therefore, the energy $H_{RKKY}$ is minimized by, e.g.,  a simple
spiral state with the wave vector $\bf q_0$ that maximizes the
non-interacting electron spin susceptibility $\chi({\bf q})$. The
momentum dependence of the susceptibility is determined by the
itinerant electron band structure. In particular, in
one-dimensional (1D) systems susceptibility diverges at twice the
Fermi momentum, and hence $q_0 = 2q_F$; in 2D continuum (low
electron density) limit it is flat up to $2q_F$, which makes all
states with $q < 2q_F$ energetically equivalent; in 3D continuum
${\bf q}_0 = 0$, which corresponds to ferromagnetism. A distortion
of the Fermi surface (and in particular ``nesting") due to lattice
discreteness effects can enhance susceptibility at a non-trivial
${\bf q}_0$, even in 3D. The main limitation of the approach based
on the quadratic Hamiltonian $H_{RKKY}$ is that it does not
energetically discriminate between the single-${\bf q}_0$ coplanar
or multiple-${\bf q}_0$ non-coplanar orderings, as long as the
constraint $|{\bf S}_i| = 1$ is satisfied \cite{footnote1}. Hence the simple description provided by
$H_{RKKY}$, while useful in determining the optimal ordering vectors
in the weak-coupling regime, is inadequate to answer the main
question of this Letter: When does the Kondo lattice model support
non-coplanar magnetic states?

To address this question we need to more accurately compare the
energy of non-coplanar magnetic states with the simpler candidates
for the ground state ordering  in the Kondo lattice model: the
ferromagnetic and the coplanar spiral states. We will focus on the
low-density 2D systems because they are expected to be
particularly prone to complex orderings, due to the massive
degeneracy within the $J^2$-order description (\ref{eq:Heff}). In
the ferromagnetic state, the free energy density at zero
temperature can be analytically evaluated as
$F_{FM}=F_0(1+J^2/\mu^2)$ when $J<\mu$ and $F_{FM}={\small
\frac{1}{2}}F_0(1+J/\mu)^2$ when $J>\mu$, where
$F_0=-(m/2\pi\hbar^2)\mu^2$ is the energy of the non-interacting
electron gas with effective mass $m$  and chemical potential $\mu=
\hbar^2q_F^2/2m$. The stability of the ferromagnetic state for $J
> \mu$ follows from the gradient expansion of the free energy,
\begin{equation}\label{eq:FF}
F - F_{FM} =  \Theta(J^2-\mu^2)\frac{J^2-\mu^2}{8\pi J}\int
\frac{d\rr}{V}|\nabla\bS|^2,
\end{equation}
where $\Theta(x) $ is the Heaviside step function and $V$ is the volume of the system.
For values of $J< \mu$, ferromagnet loses stiffness.
Remarkably, in this regime, the ferromagnetic state is {\em exactly}  degenerate (to all orders in $J$) with all spiral states, $\bS(\rr)=(\sin Ky,0,\cos Ky)$,
\begin{equation}\label{eq:FS_def}
F_S \!=\!
\sum_\pm\!\!\int\!\!\frac{d\q}{(2\pi\hbar)^2}\,[\varepsilon_\pm(\q)-\mu]\Theta(\mu-\varepsilon_\pm(\q)),
\end{equation}
\begin{equation}\label{eq:E_S}
\varepsilon_\pm(\q) = \frac{\hbar^2q^2}{2m} \!+\!
\frac{\hbar^2K^2}{8m}\pm\sqrt{\frac{\hbar^4K^2q_y^2}{4m^2}\!+\!J^2},
\end{equation}
as long as  $K \le 2q_F\sqrt{1-J/\mu}$,  see Fig. \ref{Fig:Q}a.
This degeneracy is lifted when deviations from parabolic dispersion are included. For instance, for low-density electrons hopping between nearest neighbor sites on a square lattice, there is an energetic preference towards a spiral with $K \approx 2q_F\sqrt{1-J/\mu}$, Fig. \ref{Fig:Q}b.
\begin{figure}
\includegraphics[width=0.8\columnwidth]{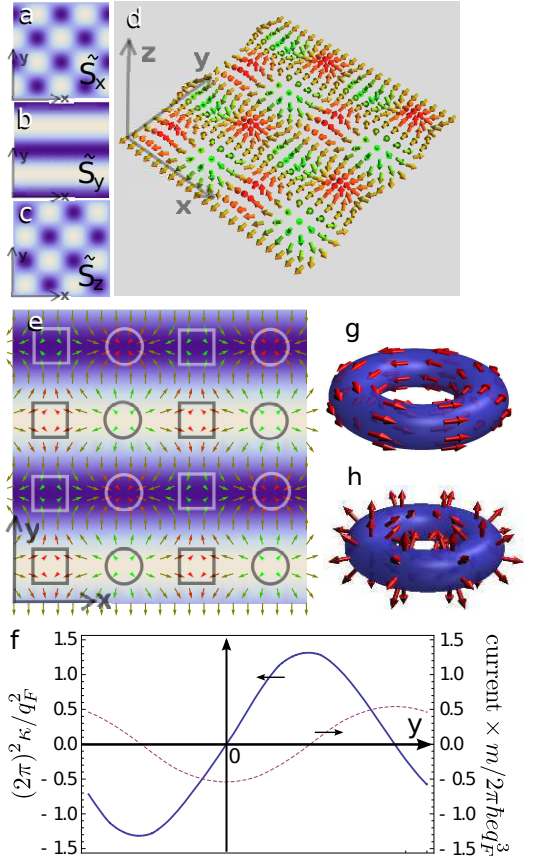}
\caption{(a-e) The optimal magnetic state $\tilde\bS(\rr) = (\sin
Ky\cos Kx,-\cos Ky,\sin Ky\sin Kx)$ realized at small Kondo
coupling, $J/\mu\ll 1$. Panels (a-c) show the spatial dependence
of the three magnetization components. (d) Full 3D magnetization
pattern. (e) The in-plane magnetization has a ``vortex-antivortex"
structure (vortex points are marked with circles, antivortex
points --- with squares). The color of the arrows (green/red)
represents the direction (down/up) of the $S_z$ component of the
local moments. The superimposed density plot in \textbf{e}
represents scalar chirality, $\kappa = {\bf S}\cdot [\partial_x
{\bf S} \times\partial_y{\bf S}]$. (f) 1D cut showing the
chirality density and the induced charge current density in the
lowest energy state, $\tilde\bS(\rr)$, for $J=0.2$ and $\mu =
0.5$. (g,h) The
two-dimensional periodic magnetization patterns can be naturally
mapped onto toroidal surface representing the real space magnetic
unit cell. A simple spiral texture, (g), is unstable with respect
to configuration (h) that corresponds to the $\tilde\bS(\rr)$
texture.}\label{Fig:tex}
\end{figure}

To determine whether the simple spiral states are unstable with respect to non-coplanar  distortions, we introduce a  slow modulation in the texture with the wave vector $Q$ perpendicular to $K$,
$\tilde\bS(\rr)=(\sin Ky\cos Qx,\sin Ky\sin Qx,\cos Ky)$. The free energy correction due to finite $Q$ can be obtained by
local spin rotation of the Hamiltonian (\ref{eq:H}) to
\begin{eqnarray}\nonumber
\hat H' &=& \frac{\hbar^2q_x^2}{2m}-\frac{\hbar^2\partial_y^2}{2m}
+ \frac{\hbar^2K^2}{8m} + \frac{\hbar^2Ki\partial_y}{2m}\sigma_y -
J\sigma_z
\\\label{eq:HUU}
&+& \frac{\hbar^2Q^2}{8m} - \frac{\hbar^2Qq_x}{2m} (\sigma_z\cos
Ky - \sigma_x\sin Ky),
\end{eqnarray}
and subsequent expansion to the second order in $Q$, which yelds
\begin{equation}\label{eq:FDS2}
F' \!= F_{FM} + [{Q^2\mu}/{8\pi}][1+I(K)] + \O(Q^4)
\end{equation}
\begin{equation}\label{eq:I_K-int}
I(K)\!=\!\! \int_{-\infty}^\infty \!\!\!\!\!\!\!\!dxdy \!\! \int_C
\!\!\!\!dz \frac{x^2\prod_\pm\!\!\left[z-2\frac{K}{q_F}\left(y\pm
\frac{K}{2q_F}\right)\right]}{\pi^2\prod_\pm\!\!\left\{z\!\!\left[z\!-\!2\frac{K}{q_F}\!\left(y\!\pm\!
\frac{K}{2q_F}\right)\right]\!\!-\!\!\frac{J^2}{\mu^2}\!\right\}}.
\end{equation}
The integration over $z$ is to be performed along the line
$\Re(z)=x^2+y^2-1$. The integrals over $z$ and $x$ can be
evaluated analytically, while the final integral over $y$ is
computed numerically. The result is presented in
Fig.~\ref{Fig:Q}c. Notably, for small $K$, the $Q^2$ correction is
identically zero. At some finite momentum
$K_0<2q_F\sqrt{1-J/\mu}$, however, the correction becomes
negative. This proves unambiguously that any simple spiral (as
well as ferromagnetic) state in the continuum limit of the Kondo
lattice model has higher energy than a more complex non-coplanar
magnetic state for $J <\mu$.  Direct numerical evaluation of the
free energy can also  be performed for arbitrary values of $Q$ and $K$, by exact diagonalization of the Hamiltonian (\ref{eq:HUU}), Fig.~\ref{Fig:Q}d.  For $J \ll \mu$, the minimum is reached by a
noncoplanar state $\tilde\bS(\rr)$ with $Q\approx K\approx q_F$.
The energetic advantage scales as ${\cal O}(J^4)$, which is
naturally outside the accuracy of the description provided by
$H_{RKKY}$.

A distinctive feature of state $\tilde S(\mathbf{r})$ is the
harmonically modulated scalar spin chirality, $\kappa \propto \sin
(Ky)$ (Fig.~\ref{Fig:tex}e), which plays a role analogous to a
spatially varying magnetic field perpendicular to the plane of the
sample. Due to the effective spin-dependent Lorentz force
acting on electrons, there are both spin and charge Hall effects
associated with the local value of $\kappa$ in the absence of an
externally applied magnetic field [charge Hall effect appears due
to the imbalance between electron spin populations aligned and
anti-aligned with the local exchange field $J \bS(\rr)$]. In the
electron band structure, magnetic ordering leads to opening of  multiple energy gaps; however, overall the system remains metallic and hence there is no quantum Hall effect. Nevertheless, similar to the quantum Hall systems where one obtains persistent edge currents, here we find
spontaneous ground state current {\em in the bulk}, with the maxima of the current magnitude occurring at the nodes of  $\kappa$, Fig \ref{Fig:tex}f.

We now discuss the possibility of other noncoplanar states. In the limit of small $J/\mu$ and very low electron density, there are stringent constraints on the kinds of states
that can be the ground state of the Hamiltonian (\ref{eq:H}). In
particular, any Fourier components $ {\bf S}_{\bf q} $ that fall
outside the circle $q < 2k_F$, would incur energy cost of order
$J^2$. For instance, this rules out the hexagonal skyrmion crystals
\cite{Rossler2006} in this regime, since for classical spins it contains an infinite number of harmonics. On the other hand, for the state $\tilde\bS(\rr)$, all Fourier harmonics ${\bf S}_{\bf q} $ can be confined to the
circle $q < 2k_F$, making it degenerate with spiral and
ferromagnetic states in the $J^2$ order, but energetically favored  in the order $J^4$. Thus, in the limit $J \ll \mu$,  ${\bf S}_{\bf q} $ is a likely candidate for the absolute ground state of the system. The highly symmetric nature of $\tilde\bS(\rr)$ can be appreciated by mapping it onto a torus representing the real space magnetic unit cell. It corresponds to the Gauss' map, which associates a normal vector to every point of the torus, Fig.~\ref{Fig:tex}h. One can easily see that the map is in the same topological sector as the spiral (Fig.~\ref{Fig:tex}g) and the ferromagnetic states (trivial map), as they can be smoothly distorted into each other without changing the real space unit cell. (This is in contrast to the map that corresponds to the hexagonal skyrmion crystal. It has a wrapping number 1, i.e. the unit sphere of $\bf S(r)$ is covered exactly once upon integration over the real space unit cell).

At intermediate values of $J/\mu$, the higher-order contributions
to energy may become comparable to the second-order (RKKY) terms
and the simple perturbative arguments can no longer be applied. In this regime,  we
speculate that a hexagonal skyrmion crystal may
become favorable as it can open more easily a full gap in the electronic
spectrum around the Fermi surface for moderate values of $J/\mu$, even though in the order
$J^2$ it loses to $\tilde\bS(\mathbf{r})$. Similarly,
quasicrystalline orders that have even higher order rotational
symmetries, e.g. 8- or 10-fold, can become competitive in the
intermediate $J/\mu$ regime since they can open the gap around the
Fermi surface even more effectively. Whether these possibilities
are realized in the model (1) is a challenging problem.

Finally, we comment on the effects of lattice discreteness, which causes
deviations from the pure parabolic dispersion assumed in the model
(\ref{eq:H}).
In Fig.~2b we have seen
that the energy of the simple spiral state decreases on a lattice. It is therefore important to analyze how the range of stability of the noncoplanar phase is modified by this effect.
By directly evaluating the energy of the spiral vs. non-coplanar state on a square lattice with only nearest neighbor hopping at finite densities
we have found that the region of
stability of the noncoplanar state behaves as $\kappa \mu<J<\mu$ with $\kappa\approx 0.4$. The value of $\kappa$ will depend on the type of tight-binding model; the better the dispersion fits the parabolic dispersion of free electrons,
the smaller the value of $\kappa$.

Even though our focus has been on the zero-temperature behavior in
2D systems, we expect that the results will remain qualitatively valid in quasi-2D systems up to finite temperatures. Consequently, one can anticipate that non-coplanar magnetism with concomitant exotic Hall behavior, may appear in a
wide range of materials and artificial structures, including
magnetic monolayers on metallic surfaces \cite{wiesen2010},
layered magnetic materials \cite{ronning2005}, dilute magnetic semiconductor films \cite{Ohno2010}, and transition
metal oxide heterostructures. In particular, recent studies of
the Hall conductivity in thin films of Mn doped GaAs show highly unusual behavior, which cannot be explained within the conventional mechanisms of the anomalous Hall effect \cite{Ohno2010}. The carrier concentrations and the exchange coupling strengths, controlled by the Mn concentration, place this system in the interesting weak-to-intermediate coupling regime considered in this Letter, indicating possible relevance of our consideration to this system.

{\acknowledgements We thank  C. D. Batista, A. Morpurgo, D.
Podolsky, R. Wiesendanger, and A. Zheludev for discussions.  This
work was carried out under the auspices of the National Nuclear
Security Administration of the U.S. Department of Energy at Los
Alamos National Laboratory under Contract No. DE-AC52-06NA25396
and supported by the LANL/LDRD Program. }



\begin{thebibliography}{10}
\expandafter\ifx\csname url\endcsname\relax
  \def\url#1{\texttt{#1}}\fi
\expandafter\ifx\csname urlprefix\endcsname\relax\def\urlprefix{URL }\fi
\providecommand{\bibinfo}[2]{#2}
\providecommand{\eprint}[2][]{\url{#2}}

\bibitem{grunberg1986}
P. Gr\"unberg, \emph{et~al.}, Phys. Rev. Lett. \textbf{57}, 2442
(1986).

\bibitem{jaime2001}
M.~B. Salamon and M. Jaime, Rev. Mod. Phys. \textbf{73}, 583
(2001).

\bibitem{Bramwell2001}
S.~T. Bramwell and M. J.~P. Gingras, Science \textbf{294}, 1495
(2001).

\bibitem{Braun2005}
H.~B. Braun, \emph{et~al.}, Nature Physics \textbf{1}, 159 (2005).

\bibitem{BodePappas}
M. Bode, \emph{et~al.}, Nature \textbf{447}, 190 (2007); C.
Pappas, \emph{et~al.}, Phys. Rev. Lett. \textbf{102}, 197202
(2009).

\bibitem{MNSIYu}
S. M{\"u}hlbauer, \emph{et~al.}, Science \textbf{323}, 915 (2009);
X.~Z. Yu, \emph{et~al.}, Nature \textbf{465}, 901 (2010).

\bibitem{Ye1999}
J. Ye, \emph{et~al.}, Phys. Rev. Lett. \textbf{83}, 3737(1999).

\bibitem{TaguchiNeubauer}
Y. Taguchi, \emph{et~al.}, Science \textbf{291}, 2573 (2001); A.
Neubauer, \emph{et~al.}, Phys. Rev. Lett. \textbf{102}, 186602
(2009).

\bibitem{Machida2010}
Y. Machida, \emph{et~al.}, Nature \textbf{463}, 210 (2009).

\bibitem{OhgushiShindou}
K. Ohgushi, S. Murakami, and N. Nagaosa, Phys. Rev. B \textbf{62},
R6065 (2000); R. Shindou and N. Nagaosa, Phys. Rev. Lett.
\textbf{87}, 116801 (2001).

\bibitem{Martin}
I. Martin and C. D. Batista, Phys. Rev. Lett. \textbf{101}, 156402
(2008).

\bibitem{Momoi1997}
T. Momoi, K. Kubo, and K. Niki, Phys. Rev. Lett. \textbf{79}, 2081
(1997).

\bibitem{domenge2008}
J.-C. Domenge, \emph{et~al.}, Phys. Rev. B \textbf{77}, 172413
(2008).

\bibitem{Chern2010}
G.-W. Chern, Phys. Rev. Lett. {\bf 105}, 226403 (2010).

\bibitem{Akagi2010}
Y. Akagi and Y. Motome, J. Phys. Soc. Jpn. {\bf 79}, 083711 (2010).

\bibitem{Kato2010}
Y. Kato, I. Martin, and C.~D. Batista, Phys. Rev. Lett. {\bf 105}, 266405 (2010) .

\bibitem{Brink2010}
S. Kumar and J. van~den Brink, Phys. Rev. Lett. \textbf{105},
216405 (2010).

\bibitem{Rossler2006}
U.~K. R{\"o}{\ss}ler, A.~N. Bogdanov, and C. Pfleiderer, Nature
\textbf{442}, 797 (2006).

\bibitem{RK1954}
M.~A. Ruderman and C. Kittel, Phys. Rev. \textbf{96}, 99 (1954).

\bibitem{Villain1977}
J. Villain, J. Phys. (Paris) \textbf{38}, 385(1977).

\bibitem{footnote1}
Thermal or
quantum fluctuations can eliminate degeneracies of classical spin
models.  In our case, we show however, that the ground state
degeneracy is naturally lifted by the more accurate treatment of
the Kondo lattice model, i.e. by going beyond quadratic
approximation.

\bibitem{wiesen2010}
S. Heinze, \emph{et~al.}, Nature Physics {\bf 7}, 713 (2011).

\bibitem{ronning2005}
F. Ronning, \emph{et~al.}, Journal of Magnetism and Magnetic
Materials \textbf{310}, 392 (2007).

\bibitem{Ohno2010}
D. Chiba, \emph{et~al.}, Phys. Rev. Lett. \textbf{104}, 106601
(2010).

\end{thebibliography}
\end{document}